\documentstyle[epsfig,12pt]{article}

\newcommand{\pom}{ $I\hspace{-1.6mm}P$}

\newcommand{\be}{\begin{equation}}
\newcommand{\ee}{\end{equation}}
\newcommand{\bea}{\begin{eqnarray}}
\newcommand{\eea}{\end{eqnarray}}
\newcommand{\bean}{\begin{eqnarray*}}
\newcommand{\eean}{\end{eqnarray*}}

\newcommand{\gapproxeq}{\lower
.7ex\hbox{$\;\stackrel{\textstyle >}{\sim}\;$}}
\newcommand{\lapproxeq}{\lower
.7ex\hbox{$\;\stackrel{\textstyle <}{\sim}\;$}}

\def\fz{$f_0(980)$}
\def\az{$a_0(980)$}

\def\fzz{$f_0(1370)$}

\def\azz{$a_0(1450)$}
\def\ss{$ s\bar s $}
\def\nn{$n\bar{n}$}

\def\qq{$q\bar q$}
\def\KK{$K\bar K$}
\def\sig{$\sigma$}
\def\lsim{\;\raise0.3ex\hbox{$<$\kern-0.75em\raise-1.1ex\hbox{$\sim$}}\;}
\def\gsim{\raise0.3ex\hbox{$>$\kern-0.75em\raise-1.1ex\hbox{$\sim$}}}
\begin{document}
\begin{titlepage}

\begin{tabbing}
right hand corner using tabbing so it looks neat and in \= \kill
\date{\today}      
\end{tabbing}
\baselineskip=18pt \vskip 0.9in
\begin{center}
{\bf \LARGE Scalar mesons above and below 1 GeV}\\
\vspace*{0.3in}
{\large Frank E. Close}\footnote{\tt{e-mail: F.Close@physics.ox.ac.uk}} \\
\vspace{.1in}
{\it Department of Theoretical Physics,
University of Oxford, \\
Keble Rd., Oxford, OX1 3NP, United Kingdom}%

\vspace{0.1in}
{\large Nils A. T\"ornqvist}\footnote{\tt{e-mail: nils.tornqvist@helsinki.fi}} \\
{\it Department of Physical Sciences},
{\it University of Helsinki,\\
PB 64 Helsinki, Finland}\\

\end{center}

\begin{abstract}
 We show that two nonets and a glueball provide a consistent description of
 data on scalar mesons below 1.7 GeV. Above 1 GeV the states form a
 conventional $q\bar{q}$ nonet mixed with the  glueball of lattice QCD. Below
 1 GeV the states also form a nonet, as implied by the attractive forces
 of QCD, but of more complicated nature. Near
 the center they are $(qq)_{\bar{3}}(\bar{q}\bar{q})_3$ in S-wave, with some $q\bar{q}$
  in P-wave, but further out they
 rearrange as $(q\bar{q})_{1}(q\bar{q})_1$ and  finally as meson-meson states.
  A simple effective chiral  model for such a
 system with two scalar nonets can be made involving two coupled
 linear sigma
 models. One of these could be looked upon as the Higgs sector of nonpertubative QCD.
\end{abstract}

\end{titlepage}

\section{The Enigmatic Scalar Mesons}

In the heavy flavour sector there are clearly established scalar mesons $c\bar{c}$ and $b\bar{b}$. They behave
as canonical $^3P_0$ states which partner $^3P_{1,2}$ siblings. Their production (e.g in radiative
transitions from $2^3S_1$ states) and decays (into $1^3S_1$ or light hadrons) are all in accord with this. There
is nothing to suggest that there is anything ``exotic" about such scalar mesons.

For light flavours too there are clearly identified $^3P_{1,2}$ nonets which call for analogous $^3P_0$ siblings.
However, while all other $J^{PC}$ combinations appear to be realised as expected, (apart from well known
and understood anomalies in the $0^{-+}$ pseudoscalars), the light scalars empirically stand out as singular.

The interpretation of the nature of the lightest scalar mesons has
been controversial for over thirty years. There is still no general agreement on
where are the  $q\bar q$ states, whether there is necessarily a glueball among the
light scalars, and whether some of the too numerous scalars are multiquark, $K\bar K$
or other meson-meson bound states.
These are fundamental questions of great importance in particle
physics. The mesons with vacuum quantum numbers are known to be
crucial for a full understanding of the symmetry breaking
mechanisms in QCD, and presumably also for confinement.

In this paper we propose a resolution of the enigma of the scalar mesons.

Theory and data are now converging that QCD forces are at work but
with different dynamics dominating below and above 1 GeV mass. The
experimental proliferation of
 light scalar
mesons is consistent with two nonets, one in the 1 GeV region  (a
meson-meson nonet) and another one near 1.5 GeV (a  $q\bar q$
nonet), with evidence for glueball degrees of freedom. At the
constituent level these arise naturally from the attractive
interquark forces of QCD. Below 1 GeV these give a strong
attraction between pairs of quarks and antiquarks in S-wave
leading to a nonet which is ``inverted" relative to the ideal
nonets of the simple $q\bar{q}$ model. Conversely, above 1 GeV the
states seeded by $^3P_0$ $q\bar{q}$ are present. The scalar
glueball, predicted by lattice QCD in the quenched approximation,
causes mixing among these states.

The phenomenon of multiplet doubling now
 requires a new effective model for the light
scalar spectrum. One possibility\cite{nils0201171} is that two coupled linear sigma models may provide
a first step for understanding such a proliferation of light scalar states.
After gauging the overall symmetry one  could then look at
the lightest scalars as Higgs-like bosons for the nonperturbative low energy
strong interactions.

Our main intuition on the strong interaction limit of QCD
derives from Lattice QCD. This impacts on scalars in three ways:

 (i) it gives a linear potential
for $q\bar{q}$ systems\cite{Ba97} which implies a nonet of scalars in the 1.2-1.6 GeV region;

(ii)  the lightest glueball, in the quenched approximation, is a
scalar with mass $\sim$1.6 GeV\cite{Ba93,Se95,fcteper,Mo97,Wein};

(iii) a strong attraction between $(qq)_{\bar{3}}$ and
$(\bar{q}\bar{q})_3$ in S-wave flavour nonet manifested below 1
GeV\cite{Jaf00,Ja77,Ja00}.

We shall argue that this guide enables us to decipher the scalar data.
We shall summarise the phenomenology that appears to be consistent with this scenario and
propose experimental tests.

We now expand on the three points above.

 (i) The predicted linear potential is well established for heavy flavours, where data confirm it,
 and where ``canonical" scalar $Q\bar{Q}$ are seen, as already mentioned. We are given a gift
of Nature in that as one comes from heavy to light flavours, the linear potential
continues phenomenologically to underpin the data: the S-P-D gaps are similar for
$b\bar{b}$, $c\bar{c}$, and even $u\bar{d}$ as can be verified by looking in the
PDG tables\cite{12gev,isgur1}. Even though we have no fundamental
understanding of why this is, we can nonetheless accept the gift and be confident that
we can assign light flavoured mesons of given $J^{PC}$ to the required ``slot" in the
spectrum\cite{Go99}. The resulting pattern leads one to expect that the lightest $J^{PC}$ $^3P_0$
$q\bar{q}$ nonet should occur in the region above 1 GeV.

There is empirically a nonet (at least) in this region above 1 GeV
and the identification of the $^3P_0$ $q\bar{q}$ nonet should be
apparent: there are candidates in $a_0(\sim
1450);f_0(1370);K(1430);f_0(1500)$ and $f_0(1710)$. One immediately
notes that if all these states are real there is an excess,
precisely as would be expected if the glueball predicted by the
lattice is mixing in this region.

(ii) Lattice QCD predictions for the mass of the lightest (scalar) glueball are now mature.
In the quenched approximation the mass is
$\sim 1.6$ GeV\cite{Ba93,Se95,fcteper,Mo97,Wein}. Flux tube models
imply that if there is a $q\bar{q}$ nonet nearby, with the same $J^{PC}$ as the glueball,
then $G-q\bar{q}$ mixing will dominate the decay\cite{Am95}.
This is found more generally\cite{An98} and recent studies on coarse-grained
lattices appear to confirm that there is indeed significant mixing between
$G$ and $q\bar{q}$
together with associated mass shifts, at least for the scalar sector\cite{McN00}.

Furthermore the maturity of the $q\bar{q}$ spectrum
tells us that we anticipate the $0^{++} q\bar{q}$ nonet to occur in the 1.2 to 1.6
GeV region. Any such states will have widths and so will mix with a scalar glueball
in the same mass range. It turns out that such mixing will lead to three physical
isoscalar states with rather characteristic flavour content\cite{fcteper,ClK01a}.
 Specifically; two will
 have the $n\bar{n}$ and $s\bar{s}$ in phase (``singlet tendency"), their mixings with
 the glueball having opposite relative phases; the third state will have the $n\bar{n}$
and $s\bar{s}$ out of phase (``octet tendency") with the glueball tending to decouple in
the limit of infinite mixing. There are now clear sightings of prominent scalar
resonances $f_0(1500)$ and $f_0(1710)$ and, probably also, $f_0(1370)$.
(Confirming the resonant status of the latter is one of the critical pieces
needed to clinch the proof - see ref.\cite{klempt} and later).  The
production and decays of these states are in remarkable agreement with this flavour
scenario\cite{ClK01a}.

A major question is whether the effects of the glueball are
localised in this region above 1 GeV, as discussed by
ref\cite{ClK01a,FC00} or spread over a wide range, perhaps down to
the $\pi\pi$ threshold\cite{minkochs}. This is the phenomenology
frontier. There are also two particular experimental issues that
need to be settled: (i) confirm the existence of $a_0(1450)$ and
determine its mass; (ii) is the $f_0(1370)$ truly resonant or is
it a $t-$channel exchange phenomenon associated with
$\rho\rho$\cite{klempt}. We return to to these resonances above 1
GeV later in the text.

Precision data on scalar meson production and decay are consistent with this and the
challenge now centres on clarifying the details and extent of such mixing.

Were this the whole story on the scalar sector there would be no doubt that the
glueball has revealed itself. However, there are features of the scalars in the
region from two pion threshold up to $O(1)$ GeV that have clouded the issue,
 in particular the existence and nature of the $f_0(980)$ and $a_0(980)$
mesons, and possibly a  $\sigma$ and a $\kappa
$ below 1 GeV.
(While the
$\sigma$ is claimed in recent data, there are conflicting
conclusions about the existence of the $\kappa$ both in
experiments\cite{kappadata}; and in
phenomenology\cite{kappa,kappaphen}).

There
 has been considerable recent progress here that enable a consistent picture to be
proposed. We set the scene for this and return later to the experimental
challenges.

(iii) QCD predicts that there is a strong attraction in S-wave between
 $(qq)^{\bar{3}}$ and $(\bar{q}\bar{q})^3$ (where superscripts
 denote their colour states) when in a
flavour nonet\cite{Ja77,Ja00}.
In addition long experience from meson-meson scattering
has shown that at low energies there is attraction only in
channels with nonexotic flavour quantum numbers, i.e. for flavour octets
and singlets. This empirical attraction between such mesons matches that of the strong attraction
between their (colour-rearranged) $(qq)^{\bar{3}}(\bar{q}\bar{q})^3$

Thus as far as the quantum numbers are concerned these
 $(qq)^{\bar{3}}(\bar{q}\bar{q})^3$ states will be like
two $0^{-+}$ $(q\bar{q})^1(q\bar{q})^1$ mesons in S-wave.
In the latter spirit, Isgur and Weinstein\cite{iswein} had
noticed that they could motivate an attraction among such mesons, to the extent
that the $f_0(980)$ and $a_0(980)$ could even be interpreted as $K\bar{K}$
molecules.

Thus the general conclusion from theory is that if there are resonances in addition to
\qq , one should expect them to form a nonet also, and that a
meson-meson component can be substantial in their wave function. Thus we anticipate that
below 1 GeV a nonet can occur with a compact $(qq)^{\bar{3}}(\bar{q}\bar{q})^3$ and a long range
meson-meson ($(q\bar{q})^1(q\bar{q})^1$)tail.

The relationship between these is being debated
\cite{tornqv,penn,achasov2,achasov1,speth}, but while the details
remain to be settled, there is a rather compelling message of the
data as follows.
 Below 1 GeV the phenomena
point clearly towards an S-wave attraction among two quarks and two antiquarks
(either as $(qq)^{\bar{3}}(\bar{q}\bar{q})^3$, or $(q\bar{q})^1(q\bar{q})^1$ ), while
above 1 GeV it is the P-wave $q\bar{q}$ that is manifested. There is a critical
distinction between them: the
``ideal" flavour pattern of a $q\bar{q}$ nonet on the one hand,
and of a $qq\bar{q}\bar{q}$ or meson-meson
 nonet on the other, are radically different; in effect they
 are flavoured inversions of one another. Thus whereas the
former has a single $s\bar{s}$ heaviest, with strange in the middle and and I=0; I=1
set lightest (``$\phi;K;\omega,\rho$-like"), the latter has the I=0; I=1
set heaviest ($K\bar{K};\pi\eta$ or $s\bar{s}(u\bar{u} \pm d\bar{d}$)) with strange in the
middle and an isolated I=0 lightest ($\pi\pi$ or $u\bar{u}d\bar{d}$)\cite{Jaf00,Ja77,iswein}.

The phenomenology of the $0^{++}$ sector appears to exhibit both of these patterns with
$\sim 1$GeV being the critical threshold. Below 1 GeV the inverted structure of the four quark
dynamics in S-wave is revealed with $f_0(980);a_0(980)$; $\kappa$ and $\sigma$ as the labels.
One can debate whether these are truly resonant or instead are
the effects of attractive long-range $t-$channel dynamics
 between
the colour singlet $0^{-+}$ $K\bar{K}; K\pi; \pi\pi$, i.e.,
whether they are meson-meson molecules or $qq\bar{q}\bar{q}$. But
the systematics of the underlying dynamics seems clear.

The phenomena are consistent with a strong attraction of QCD in
the scalar S-wave nonet channels. The difference between molecules
and compact $qq\bar{q}\bar{q}$ will be revealed (provided phase
space effects are removed) in the
tendency for the former to decay into a single dominant channel -
the molecular constituents - while the latter will feed a range of
channels driven by the flavour spin clebsch gordans. For the light
scalars it has its analogue in the production characteristics.

The picture that is now emerging from both
phenomenology\cite{ClK01b,ClK01c,kloe} and the\-ory\cite{jaffe01}
is that both components are present. As concerns the
theory\cite{jaffe01}, think for example of the two component
picture as two channels. One, the quarkish channel ($QQ$) is
somehow associated with the $(qq)^{\bar 3}(\bar{q}\bar{q})^{3}$
coupling of a two quark-two antiquark system, and is where the
attraction comes from. The other, the meson-meson channel ($MM$)
could  even  be completely passive (eg, no potential at all).
There is some off diagonal potential which flips that system from
the $QQ$ channel to $MM$. The way the object appears to experiment
depends on the strength of the attraction in the $QQ$ channel and
the strength of the off-diagonal potential. The nearness of the
$f_0$ and $a_0$ to $K\bar{K}$ threshold suggests that the $QQ$
component cannot be too dominant, but the fact that there is an
attraction at all means that the $QQ$ component cannot be
negligible. So in this line of argument, $a_0$ and $f_0$ must be
superpositions of four-quark states and $K\bar{K}$ molecules.

 Continuing this
argument to include a coupling to \qq\ they can be superpositions of
all 3 configurations (\qq , 4-quark, and meson-meson). However, for a scalar the $q\bar{q}$ system is in P-wave
and naturally higher in energy. The $qq\bar{q}\bar{q}$ is strongly attracted in S-wave
with no angular momentum barrier. Thus in a
spatial picture one expects the four-quark S-wave state to form
the core, while the outer regime (which extends to distances
inversly proportional to $2m_K-m_{res}$) is composed of \KK, with any residual \qq\ P-wave at intermediate range.

\section{Heavy Flavours and Light Scalars}

 The working hypothesis is that the $q\bar{q}$ nonet, mixed with a glueball, is realised
above 1 GeV and we now need to determine the
flavour content of these states. In addition we need
to confirm the picture of the $f_0(980)$ and
light scalars below 1 GeV. New opportunities for improving data are coming from
heavy flavour decays, in particular $D_s$ decays.

\subsection{Scalars $\geq$ 1 GeV}

The decay $\psi \to \gamma \pi\pi$ compared with $D_s \to \pi \pi \pi$, and
$\psi \to \gamma K\bar{K}$ compared with $D_s \to \pi K\bar{K}$
provide complementary entrees into the light flavoured $0^{++}$ mesons.  Comparison with
$D_d \to \pi K^*_0(1430)$ then enables
us to ``weigh" the flavour content of the nonets. In
$\psi \to \gamma K\bar{K}$ Dunwoodie\cite{dunwoodie} finds the $f_0(1710)$ as clear
scalar, and this state could be sought in $D_s \to \pi K\bar{K}$
with enough statistics (in E687\cite{e687}
the $K^*\bar{K}$ band contaminates the 1710 region of the Dalitz plot). This
could be a challenge for high statistics data e.g. with
FOCUS. The major signal in the E687 data is
the $\phi$; the $f_0(980)$ is just below threshold and it is not discussed whether
any of the signal at threshold is due to this state. However, in the E791 data\cite{e791} on
$D_s \to \pi \pi \pi$ the $f_0(980)$ is very prominent (see next subsection), together with the $f_0(1370)$
and a possible (though unclaimed) hint of a shoulder that could
signal the $f_0(1500)$. Dunwoodie's analysis of $\psi \to \gamma \pi\pi$
shows structure around 1400 MeV and with better statistics from BES and
Cornell this should be verified and attempts made to resolve it into $f_0(1370)$
and $f_0(1500)$. The strength of $f_0(1710)$ in these data should also
be determined.

\subsection{Scalars below 1 GeV}

In the 22
different analyses on the \sig\ pole position, which are included
in the  2000 edition of the Review of Particle Physics
\cite{pdg2000} under the entry $f_0(400-1200)$ or \sig,
 most  find a \sig\ pole position near 500-i250 MeV.
 Also, at  a recent meeting in Kyoto \cite{kyoto} devoted to the
$\sigma$, many groups reported preliminary analyses, which find
 \sig\ resonance parameters in the same region. Furthermore as
we discuss in more detail below the \sig\ has been claimed in
$D\to 3\pi$ and  in $\tau$ decay. There is also a very clear,
although still preliminary, signal for a light \sig\ (Breit-Wigner
mass=$390^{+60}_{-36}$ and width= $282^{+77}_{-50}$) in a BES
experiment\cite{wu} on
J/$\psi\to\sigma\omega\to\pi\pi\omega$.

Barnes\cite{barnes02} has warned that it may be premature to infer
 resonance
parameters from such data. Only the low energy tails of the
purported resonance phase shifts are actually in evidence in the
charm data and crucial observation of a complete Breit-Wigner
phase motion through 180$^o$ has not been made. Especially the
elastic $K\pi$ phase shift as measured by LASS\cite{aston} (but
also the
 $\pi\pi$ phase shift in many models) does not show evidence of a ``complete" low mass scalar
resonance. Therefore concluding that these resonances exist based
on the charm data in isolation, which only covers part of the
range of invariant mass that has already been studied in light
hadronic processes, is unjustified. Barnes recommends that the
charm decay analyses should include what is already known about
phase shift analyses over the full mass range, e.g. from
ref\cite{au}.

In this connection it should be remembered that
Breit-Wigner parameters and pole positions,
 can differ by several 100 MeV for the same data.
 (Also mass parameters which appear in phenomenological
 Lagrangian similarily  differ considerably from these masses.)
Therefore the uncertainty in mass determinations of a broad
resonance like the \sig, or the $\kappa$ is not only due to
experimental uncertainties, but also depend on  the definitions of
mass used.

An interesting piece of data also comes  from the CLEO analysis of
$\tau\to a_1\nu \to \sigma\pi\nu\to 3\pi\nu$ \cite{CLEO}, which
finds a \sig\ BW mass of approximately 555 MeV and a width of 540
MeV. Perhaps more importantly, their branching ratio for $a_1$ of
$ \Gamma_{\sigma\pi}/\Gamma_{tot} =.16$ should make S. Weinberg
happy, since he urged people to look for this decay mode, and
predicted a 50 MeV partial decay mode for $a_1\to \sigma \pi$ in
his ``mended symmetry" paper of 1990\cite{wein90}, at a time when
the $a_1\to \pi(\pi\pi)_{S-wave}$ was quoted to be essentially
absent (0.7\%). Although $\Gamma_{tot}(a_1)$ is very uncertain,
250-600 MeV, this agrees well with Weinberg's estimate.

The recent experiments studying charm decay to light hadrons are
opening up a new experimental window for understanding light meson
spectroscopy and especially the controversial scalar mesons, which
are copiously produced in these decays. We therefore discuss in
more detail the recently measured $D\to\sigma\pi\to 3\pi$ and
$D_s\to f_0(980)\pi\to 3\pi$ decays, where the \sig , respectively
the $f_0(980)$, is clearly seen as the dominant peak, and point
out that these decays rates can be understood in a rather general
model for the weak matrix elements. This indicates that the broad
\sig (600) and the $f_0(980)$ belong to the same multiplet.

In particular we refer to the E791 study of the $D\to 3\pi$ decay
\cite{e791} where it is shown how adding an intermediate
S-wave structure with floating mass and width in the Monte Carlo program
simulating the Dalitz plot densities, allows for an excellent fit
to data provided the mass and the width of this scalar
are $m_\sigma\simeq 478$ MeV and $\Gamma_\sigma\simeq 324$ MeV.
In fact 46\% of the $D^+\to3\pi$ Dalitz plot was explained by their $\sigma\pi$.

In a calculation by Gatto et al.\cite{gatto}
this hypothesis was checked  adopting the E791 experimental values for its mass
and width and using a Constituent Quark Meson Model (CQM) for
heavy-light meson decays \cite{rass}. The $D\to\sigma\pi$ non-leptonic process
was computed, assuming
 {\it factorization}
\cite{WSB}, and taking the coupling of the $\sigma$ to the light
quarks from the linear sigma model \cite{volkoff}. In such a way
one is directly assuming that the scalar state needed in the E791
analysis could be the quantum of the $\sigma$ field of the linear
sigma model. According to the CQM model and to factorization, the
amplitude describing the $D\to\sigma\pi$ decay can be written as a
product of the semileptonic amplitude $\langle
\sigma|A^\mu_{(\bar{d}c)}(q)|D^+\rangle$, where $A^\mu$ is the
axial quark current, and $\langle \pi|A_{\mu(\bar{u}d)}(q)|{\rm
VAC}\rangle$. This computation indicates that the low mass S-wave
enhancement described in the E791 paper can be consistently
understood as the $\sigma$ of the linear sigma model. In a rather
similar approach Paver and Riazuddin\cite{paver} also reach the
same conclusion.

These models\cite{paver,deandrea} were also used to predict the
related process $D_s\to f_0(980)\pi\to 3\pi$. The agreement with
the data, where the $f_0(980)$ is dominant in the Dalitz plot,
indicates that the $\sigma$ and the $f_0(980)$ belong to a similar
flavour multiplet. Also when comparing the
predicted rates of \sig\ and $f_0(980)$ in these charm decays to
those predicted for $\rho$ and $\phi(1020)$ using the same models
one finds a good agreement with the data.

Note however as we discussed above \cite{barnes02} that it may be
premature to infer broad \sig\ and $\kappa$ resonance parameters
from such data. Only the low energy tails of the purported
resonance phase shifts are actually in evidence in the charm data
and crucial observation of a complete Breit-Wigner phase motion
through 180$^o$ has not been made.

But, whatever the true nature of this structure, there is clear
evidence for a strong attractive enhancement at low $m(\pi\pi)$ in
S-wave which we can denote by $\sigma$. Its dynamical origin and
parameters remain to be settled. Further theoretical studies are
now needed to propose alternative explanations of the E791 data.
These are required to enable useful comparison of points of view
on the nature of the $\sigma$.

One clear message from the E791 data is that $f_0(980)$ has strong
affinity for $s\bar{s}$ in its production at short distances. Then
it evolves building up a substantial virtual \KK\ component at
larger distances, and decays violating the OZI rule in a two step
process, into $\pi\pi$ which is the only open channel.

There is a large amount of data on the production of the
$f_0(980)$ which require in some cases a strong affinity for
$s\bar{s}$ (e.g. the $D_s$ decays mentioned above), or for $n\bar{n}$ (the
production in hadronic Z decays has all the characteristics
associated with well established $n\bar{n}$
states\cite{lafferty,delphi}) and also data that require both
components to be present ($\psi \to  \omega f_0$ versus $\psi \to
\phi f_0$). There are also new data from central production and on $\phi \to \gamma f_0(a_0)$
that touch on the relationship
between $f_0-a_0$. We shall now discuss these.

\section{The light Scalars $f_0/a_0(980)$: Phenomena and Theory}

\subsection{$f_0$ $a_0$ mixing in central production}

Further evidence that the dynamics of $f_0-a_0(980)$ are strongly influenced by the $K\bar{K}$ threshold
(see \cite{ach81} for an early theoretical discussion and estimate of this effect) is
the presence of strong mixing and violation of isospin or G-parity for these states.
A more detailed review of ideas on $\phi \to \gamma f_0/\gamma a_0$ and
 the implications of the mixing hypothesis on these data is in ref.\cite{fecrome}.
Emerging data from DA$\Phi$NE
are in remarkable agreement with these predictions and add weight to
the idea that these scalar mesons are compact $qq\bar{q}\bar{q}$ states with
an extended meson-meson cloud ``molecular" tail.

The $x_F$ distribution for the
$a_0^0(980)$ in $p p \to p p a_{0,2}$ is the only state with $I=1$ that is observed
to have a $x_F$ distribution peaked at zero~\cite{sumpap},
and moreover the distribution for the $a_0^0(980)$
looks similar to  the central production of states that are accessible to
\pom \pom \thinspace fusion,
in particular \pom \pom $\rightarrow f_0(980)$

The $\phi$ distribution for the $a_0(980)$ also looks very similar
to that observed for the $f_0(980)$. Qualitatively this is what
would be expected if part of the centrally produced $a_0^0(980)$
is due to \pom \pom $\to f_0(980)$ followed by mixing between the
$f_0(980)$ and the $a_0(980)$.
\par
Ref \cite{ClK01b} found
 that 80~$\pm$~25~\% of the $a_0^0(980)$ comes from the
$f_0(980)$ and upon combining this result with the relative total
cross sections for the production of the $f_0(980)$ and
$a_0^0(980)$~\cite{sumpap} they found the $f_0(980)-a_0(980)$
mixing intensity to be 8~$\pm$~3~\%.
Achasov et al\cite{ach81} predicted an order of magnitude
$0.5-2$\% for this mixing based upon the affinity of the scalars
for the nearby $K\bar{K}$ threshold. The data appear larger even
than this and
 add weight to the hypothesis that the
$f_0(980)$ and $a_0(980)$
are siblings that strongly mix, and that the $a_0(980)$ is not simply
a $^3P_0$ $q\bar{q}$ partner of
the $a_2(1320)$. This is consistent with the $0^{++}(QQ/MM)$ picture
of these states and
a natural explanation of these results is that
$K\bar{K}$ threshold plays an essential role in the existence and
properties.
\par
Other lines of study are now warranted.
Experimentally to confirm these ideas requires measuring the
production of the $\eta\pi$ channel at a much higher energy, for
example, at LHC, Fermilab
or RHIC where any residual Reggeon exchanges such as $\rho \omega$
would be effectively zero
and hence any $a_0(980)$ production must come from isospin breaking
effects.
\par
Other ``pure" flavour channels should now be
explored. Examples are $D_s$ decays
where the weak decay leads to a pure I=1 light hadron final state. Thus
$\pi f_0(980)$ will be
(and is~\cite{e791}) prominent, while the mixing results suggest
that $\pi a_0$ should also be present at $8 \pm 3$ \% intensity. Studies
 with high statistics data sets now emerging
from E791, Focus and BaBar are called for, and also
 studies of $J/\psi$ decays at Beijing,
 in particular to the
``forbidden" final states $\omega a_0$ and $\phi a_0$ where ref\cite{ClK01b}
 predicts
branching ratios of $O(10^{-5})$. Decays of $f_1(1285) \to \pi\pi\eta$ should also be
accompanied by  $f_1(1285) \to \pi\pi\pi$ \cite{ach81}.

\subsection{$\phi \to \gamma f_0/\gamma a_0$}

The radiative decays of the $\phi \to \gamma f_0(980)$ and
$\gamma a_0(980)$ have long been recognised as a potential route towards disentangling
their nature.
 Isospin mixing effects could considerably alter
some predictions in the literature for $\Gamma(\phi \to \gamma f_0(980))$ and
$\Gamma(\phi \to  \gamma a_0(980))$, and new data from DA$\Phi$NE
promise to reveal their nature.

The magnitudes of these widths are predicted to be rather
sensitive to the fundamental structures
of the $f_0$ and $a_0$, and as such potentially discriminate amongst them.
For example, if $f_0(980) \equiv s\bar{s}$ and the dominant dynamics is
the ``direct" quark transition
$\phi(s\bar{s}) \to \gamma 0^{++}(s\bar{s})$, then
 the predicted $b.r.(\phi \to \gamma f_0)
\sim 10^{-5}$, the rate to $\phi \to \gamma a_0(q\bar{q})$ being even
smaller due to OZI supression\cite{cik}.
 For $K\bar{K}$ molecules the rate was predicted
to be higher, $\sim (0.4 - 1) \times 10^{-4}$\cite{cik}, while for tightly compact
$qq\bar{q}\bar{q}$ states the rate is yet higher, $\sim 2 \times 10^{-4}$\cite{cik,achasov3}.

In the $K\bar{K}$ molecule and $qq\bar{q}\bar{q}$ scenarios
it has uniformly been assumed that the radiative transition will be
driven by an intermediate  $K^+K^-$ loop ($\phi \to K^+K^-
\to \gamma K^+K^-  \to \gamma 0^{++}$). Explicit calculations
in the literature agree that this implies\cite{cik,achasov3,cbrown,lucieu}

\begin{equation}
b.r. (\phi \to f_0(980)\gamma ) \sim (2 \pm 0.5) (10^{-4}) \times
F^2(R)
\end{equation}

\noindent where $F^2(R) = 1$ in point-like effective field theory
computations, such as refs.\cite{achasov3,lucieu}.
By contrast, if the $f_0(980)$ and $ a_0(980)$ are spatially extended $K\bar{K}$
molecules, (with r.m.s. radius $R > O(\Lambda_{QCD}^{-1}$)),
then  the high momentum region of the integration
in  refs.\cite{cik,cbrown} is cut off, leading in effect
to a form factor suppression, $F^2(R) < 1$\cite{cik,achasov4,barnes}.
The differences in absolute rates are thus intimately linked to the
model dependent magnitude of $F^2(R)$.

If $f_0$ and $a_0^0$ have common constituents (and hence are ``siblings")
and are eigenstates of isospin, then their
affinity for $K^+K^-$ should be the same and so\cite{cik,achasov3,lucieu}

\begin{equation}
\frac{\Gamma(\phi \to  f_0 \gamma)}{\Gamma(\phi \to  a_0 \gamma)} \sim 1
\end{equation}
\noindent whereas the preliminary data find\cite{kloe}

\begin{equation}
\frac{\Gamma(\phi \to \gamma f_0)}{\Gamma(\phi \to \gamma a_0)}
= 4.1 \pm 0.4
\end{equation}

\noindent or even, in their more recent report, find\cite{kloe2}

\begin{equation}
\frac{\Gamma(\phi \to \gamma f_0)}{\Gamma(\phi \to \gamma a_0)}
= 6.1 \pm 0.6
\end{equation}

On this ratio alone one might conclude evidence that the states are \qq~ and that the $a_0$ is relatively
suppressed due to its $u\bar{u}-d\bar{d}$ content. However, this is not compatible with the intrinsically
``large" branching ratios of $\geq O(10^{-4})$.

Ref\cite{ClK01c}  noted that the $\eta \pi$ signal in central production
above, if described by an isospin mixing angle $\theta$, would lead to
the relative rates for $phi$ radiative to be
\begin{equation}
\frac{\Gamma(\phi \to \gamma f_0)}{\Gamma(\phi \to \gamma a_0)} \sim \frac{g^2_{f_0K^+K^-}}{g^2_{a_0K^+K^-}}
\equiv cot^2\theta
= 3.2 \pm 0.8
\end{equation}
\noindent It is intriguing that such a correlation appears to be realised. A problem
though, as emphasised by ref.\cite{ach2001} is that the significant overlap of
$f_0$ and $a_0$ states severely reduces such effects, such that one would expect the
radiative ratio of unity to survive.

 In order to use the individual rates to abstract magnitudes
of $F^2(R)$, and hence assess how compact the four-quark
state is, a definitive accurate value for $g_{fKK}^2/4\pi $ will be required.
If for orientation we adhere to the value used elsewhere, $g_{fKK}^2/4\pi\sim 0.6$
GeV$^2$,
 and impose the measured ratio above as an effective measure of the relative
couplings to the $K^+K^-$ intermediate state,
then the results of ref.~\cite{cik} are revised to

\begin{equation}
b.r.(\phi \to \gamma f_0) + b.r.(\phi \to \gamma a_0) \leq (4 \pm 1) (10^{-4})
\end{equation}

\noindent and

\begin{equation}
b.r.(\phi \to \gamma f_0) =  (3.0 \pm 0.6)  10^{-4} F^2(R)
\end{equation}
\begin{equation}
b.r.(\phi \to \gamma a_0) =  (1.0 \pm 0.25) 10^{-4} F^2(R)
\end{equation}

 Branching ratios for
which $F^2(R) << 1$ would imply that the $K^+K^-0^{++}$
interaction is spatially extended, $R > O(\Lambda_{QCD}^{-1})$. Conversely,
for $F^2(R) \to 1$, the system would be spatially compact, as in $qq\bar{q}\bar{q}$.

 The preliminary data from KLOE are \cite{kloe}

\begin{equation}
b.r.(\phi \to \gamma f_0) =  (2.4 \pm 0.1)  10^{-4}
\end{equation}
\begin{equation}
b.r.(\phi \to \gamma a_0) =  (0.6 \pm 0.05) 10^{-4}
\end{equation}

\noindent which imply $F^2(R) \sim 0.7 \pm 0.2$, supporting
the qualitative picture  of a compact $qq\bar{q}\bar{q}$ structure
that spends a sizeable part of its lifetime at longer range in a two meson state, such
as $K\bar{K}$. Subsequently they have reported
\cite{kloe2}

\begin{equation}
b.r.(\phi \to \gamma f_0) =  (4.5 \pm 0.2)  10^{-4}
\end{equation}
\begin{equation}
b.r.(\phi \to \gamma a_0) =  (0.75 \pm 0.07) 10^{-4}
\end{equation}

Caution is needed before over-interpreting these numbers. There are discrepancies between
the magnitude of $S-K\bar{K}$ couplings in this experiment and some others. This may be
connected to the fact that no $\pi\pi-KK$ coupled-channel analysis incorporating unitarity
has yet been performed on
the data. Also, the results depend upon the assumption that there is a large destructive
interference in the data between the $f_0\gamma$ and $\sigma \gamma$, where the $\sigma$ is assumed to
be described by a simple Breit-Wigner shape. The sensitivity to this assumption has not been
discussed and needs to be assessed before strong conclusions about the $f_0 \gamma$ branching ratio
are made. Hence we delay detailed interpretation pending a more detailed analysis of these data.
Certainly this process offers potentially significant insights into the nature of the scalar
enhancements below 1GeV.

\section{ Understanding the S-waves within a unitarized quark model (UQM)}

\subsection{The Nambu-Jona-Lasinio (NJL) model and the linear sigma model}

A light scalar-isoscalar meson (the \sig ), with a mass of twice
the constituent  $u,d$ quark mass, or $\approx 600$ MeV, coupling
strongly to $\pi\pi$ is of importance in all
Nambu--Jona-Lasinio-like (NJL-like) models for dynamical breaking
of chiral symmetry. In these models the $\sigma$ field obtains a
vacuum expectation value, i.e., one has a \sig-like
condensate in the vacuum, which is crucial for the understanding
of all hadron masses, as it explains in a simple way the
difference between the light constituent and chiral quark mass.
 Then most of the nucleon mass is generated by its coupling to
the $\sigma$, which acts like an effective Higgs-like
 boson for the hadron spectrum.

The NJL model is an effective theory which is believed to be
related to QCD at low energies, when one has integrated out the
gluon fields. It involves a linear realization of chiral symmetry.
After bosonization of the NJL model one finds essentially the
linear sigma model  as an approximate effective theory for
the scalar and pseudoscalar meson sector.

About  30 years ago Schechter and Ueda\cite{u3u3}  wrote down the
$U3\times U3$ linear sigma model (broken by 2 quark mass terms and
a $U_{1A}$ term) for the meson sector involving a scalar and a
pseudoscalar nonet. This (renormalizable) theory has only 6
parameters, out of which 5 can be fixed by the pseudoscalar masses
and decay constants; $m_\pi,\ m_K, \ f_\pi, \ f_K$, and a
combination of $\eta$ and $\eta'$ masses (like
$m_\eta^2+m^2_{\eta'}$), which fixes the strength of the $U_{1A}$
breaking term.

The sixth parameter for the OZI rule violating 4-point coupling is
small, and fixes the \sig\ - $a_0$ splitting. One can then
predict, with no free parameters, the other tree level scalar
masses \cite{lsm}, which turn out to be not far from the lightest
experimental masses, although the two quantities (say Lagrangian
mass vs. second sheet pole mass) are \underline{not} the same
thing, but can differ for the same model and data by well over 100
MeV.

The important thing is that the scalar masses are predicted to be
near the lightest experimentally seen scalar masses, and not in
the $\sim$1.2-1.6 GeV region where we expect  the lightest $q\bar
q$ scalars. The \sig\ is predicted \cite{lsm} at 620 MeV with  a
very large width ($\approx 600$ MeV) which agrees well with most
data. The $a_0(980)$ is predicted at 1128 MeV,  the $f_0(980)$ at
1190 MeV, and the $\kappa$ or $K^*_0(1430)$ at 1120 MeV. This is
still surprisingly good considering that loops or unitarity
effects must be large as we discuss next.

\subsection{Unitarisation and S-waves in  quark models}

A few years ago one of us presented fits to the $K\pi$, $\pi\pi$ S
-waves and to the  \az\ resonance peak in $\pi\eta$\cite{NAT1}. A
similar model was also  presented by Van Beveren et
al.\cite{beve}. This involved, like the linear sigma model, the
pseudoscalar nonet (taken from data) and a scalar nonet. It may be
looked upon as a way of unitarising a chiral quark model for the
scalars, and in a coupled channel framework it takes account of
all the flavour related s-channel two pseudoscalar thresholds.
These are shown to distort any simple input bare scalar spectrum,
through the mixing with the meson-meson continuum.

Also in this approach one must make simplifications; e.g. one neglects more distant singularities and assumes that
crossed channel singularities can be represented by a simple form factor.
But, a nice feature of such a model is that it simultaneously describes a  whole scalar
nonet and gives a good representation of a large
set of relevant data, with a few physically well defined parameters.

Consistency with unitarity implies that when the effective
coupling becomes large enough, twice as many poles can appear in
the output spectrum as were put in  as bare nonet masses. The new
poles can then be interpreted as being mainly meson-meson bound
states, but mixed with the states, which are put in.

At first\cite{tornqv} the \sig\ was missed because only poles
nearest to the physical region were looked for, and the
possibility of the resonance doubling phenomenon, discussed below,
was overlooked. Only later was it was realised \cite{NAT1} that
two resonances can emerge although only one  bare state is put in,
i.e., one bare nonet can give rise to two nonets in the output, if
the overall coupling is large enough.
 Then one had to look deeper into the second sheet and the broad \sig\ as the
dominant singularity at low mass was found in the model.
 An advantage with this model was that in
order to explore whether this pole was the relevant one, one
could, within the model, decouple the effect of the \KK\ threshold
and find that the relevant singularity in the $n\bar n$ channel is
indeed the broad \sig .

In fact, it had been pointed out by Morgan and
Pennington~\cite{morgan} that for each \qq\ state there are, in
general, apart from the nearest pole, also  image poles, usually
located far from the physical region. As explained in more detail
in Ref.~\cite{NAT1,bogl}, some of these can (for a large enough
coupling and sufficiently heavy threshold) come so close to the
physical region that they make new resonances. And, in fact, there
were more than four physical poles  with different isospin, in the
output spectrum of the UQM model, although only four bare states,
of {\it the same nonet},  were put in!.

In the I=1 channel two manifestations of the bare state were
found, the \az\ and the  \azz . Similarily for one input bare \ss\
state, two poles the \fz\ and a heavier one, which at the time was
assumed to be the \fzz\ were found, but the uncertainties of the
model could in reality push the heavier state up in mass, e.g. to
emerge within the $f_0(1500/1710)$ states. In the $K\pi$ channel a
stronger overall coupling could within the model produce a virtual
bound state  $K\pi$ state  near the threshold\cite{NAT1}.

Cherry and Pennington \cite{cherry}
 have strongly argued against the existence of a light $\kappa$.
With the presently known  experimental $K\pi$ phase
shifts\cite{aston}, and the fact that it is essentially a single
channel problem up to the $K^*_0(1430)$, this conclusion seems
hard to avoid. But it should however, be
remembered that the experimental phase shifts start at only at
about $\sqrt s\approx$850 MeV. New data on these phase shifts
would be very welcome. As we have mentioned the E791 experiment
see some evidence for  a light $\kappa$ in $D^+\to K^-\pi^+\pi^+$.
The signal is much less evident than the $\sigma$ in $D\to 3\pi$,
but the $\kappa$ improves their $\chi^2$ in the region dominated
by the $K^*(890)$.

 There are several  authors\cite{beve,black,
 modelskappa},
who within models support the existence of both the \sig\ and the
$\kappa$, but the question of whether the $\kappa$ is truly
resonant or whether it could be something like a virtual bound
$K\pi$ state remains open, and the fact that there is activity
with these quantum numbers appears to be established.

Although the details of any modelling can be criticized, the
conclusion remains: {after unitarisation, \it strong enough
couplings can generate new bound states or resonances that were
not present in the input or in the Born terms represented by an
effective Lagrangian}. In short, in addition to a conventional
scalar nonet the unitarisation  generates for large effective
coupling another scalar nonet, which has a substantial component
of meson-meson in its wave function. A more detailed QCD inspired
UQM-like model, with better description of  crossed channels, and
more thresholds would be very welcome.

There are other effects that a model like the UQM explains, which
are due to the fact that the inverse meson propagator is not just
BW-like ($m_0^2 -s +ig^2{\rm Im}\Pi(s)$), with constant $m_0$, but
has an important cusp-like contribution $m_0^2 \to m_0^2+g^2{\rm
Re}\Pi(s)$ to the mass term. For resonances decaying in an S-wave
near a threshold and with a large $g^2$ this makes a big
difference in the mass, width and shape of the resonance. In many
unitarisation schemes (e.g. often in K-matrix unitarisation) this
requirement from analyticity is forgotten. We list here the most
important effects.

 (i) {\it The large mass difference between the $K^*_0(1430)$
and the $a_0(980)$}.

 This arises as a secondary effect due to the
large pseudoscalar mass splittings, and because of the large mass
shifts coming from the loop diagrams involving the PP thresholds.
The three thresholds $\pi\eta,\ K\bar K,\ \pi\eta'$ all lie
relatively close to the $a_0(980)$.  All three of them contribute
to a large negative shift in mass, and to a large meson-meson
component in $a_0(980)$, mainly \KK . On the other hand, for the
$K^*_0(1430)$, the $SU3_f$ related thresholds ($K\pi,\ K\eta'$)
lie far apart from the $K^*_0$, while the $K\eta$ nearly decouples
because of the physical value of the pseudoscalar mixing angle.
Therefore the $K^*_0(1430)$ is shifted down only slightly and
furthermore remains essentially as \qq .
Conversely, the $\kappa$ would be dominantly $K\pi$ (provided it
forms a bound state near $K\pi$) and the $a_0(1430)$ is dominantly
$u\bar{d}$.

(ii) {\it The nearness of  the $a_0(980)$ and the $f_0(980)$  to the
\KK\ threshold.}

 Because of the cusp in $m_0^2+g^2{\rm Re}\Pi(s)$
at a threshold, a physical mass emerges just below the threshold
for a wide range of $m_0^2$ values.

(iii) {\it The narrowness of  the $a_0(980)$/$f_0(980)$ peak width.}

 The
cusp also changes the shape of the resonance peak to be
considerably narrower than what a BW parameterization gives.
This is the Flatt\'{e}\cite{flatte}
effect. In a space-like picture it is related to the fact that the
large \KK\ component in the wave function must convert near the
origin to $\pi\eta$ or $\pi\pi$, which is the only  open channel.

We now turn to the question of the scalar glueball. This will mix
with the other scalars, whether below or above 1 GeV mass. The
first issue therefore is how one might isolate such a state from
data.

\section{Glueball production dynamics}

The folklore has been that to enhance glueball signals one should concentrate on
production mechanisms where quarks are disfavoured: thus $\psi \to \gamma G$\cite{chan},
$p\bar{p} \to \pi + G$ in annihilation at rest\cite{chan,fcrpp}, and central production
in diffractive (gluonic pomeron) processes, $pp \to p G p$\cite{fcrpp,robson}.
Contrasting this,
$\gamma \gamma$ production should favour flavoured states such as $q\bar{q}$.
Thus observing a state in the first three, which is absent in the latter,
would be prima facie evidence.

Such ideas are simplistic. There has been progress in quantifying them
and in the associated phenomenology. The central production has matured
significantly in the last three years and inspires new experiments
at RHIC, Fermilab and possibly even the LHC. These complementary processes collectively are
now painting a clearer picture.

First on the theoretical front, each of these has threats and opportunities.
(i)In $\psi \to \gamma G$ the gluons are timelike and so it is reasonable
to suppose that glueball will be favoured over $q\bar{q}$ production.
Quantification of this has been discussed in ref.\cite{cfl} with some
tantalising implications: (a) the $f_0(1500)$ and $f_0(1710)$ are produced with
strengths consistent with them being $G-q\bar{q}$ mixtures, though there are
some inconsistencies between data sets that need to be settled
experimentally.

(ii) In $p\bar{p}$ the $q$ and $\bar{q}$ can rearrange themselves to produce
mesons without need for annihilation. So although a light glueball may
be produced, it will be in competition with conventional mesons and
any mixed state will be produced significantly by its
$q\bar{q}$ components.

(iii) In central production the gluons are spacelike and so must rescatter
in order to produce either a glueball or $q\bar{q}$. Thus here again one
expects competition. However, a kinematic filter has been discovered\cite{ck97}, which
appears able to suppress established $q\bar{q}$ states, when the $q\bar{q}$
are in P and higher waves.

Its essence was that the pattern of resonances produced in the central
region of double tagged $pp \rightarrow pMp$ depends on the vector
$ difference$ of the transverse momentum recoil of the final state
protons (even at fixed four momentum transfers). When this quantity
($dP_T \equiv |\vec{k_{T1}} - \vec{k_{T2}}|$) is ``large", ($\geq O(\Lambda
_{QCD})$), $q\bar{q}$ states are prominent whereas at ``small"
$dP_T$ ($\leq O(\Lambda
_{QCD})$)all well established $q\bar{q}$ are observed to be suppressed
while the surviving resonances include the enigmatic $f_0(1500),
f_0(1710)$ and $f_0(980)$.

The data are consistent with the hypothesis that as $dP_T \rightarrow 0$
all bound states with internal $L > 0$ (e.g. $^3P_{0,2}$ $q\bar{q}$)
are suppressed while S-waves survive (e.g. $0^{++}$ or $2^{++}$ glueball
made of vector gluons and the $f_0(980)$ as any of glueball,
or S-wave $qq\bar{q}\bar{q}$ or $K \bar{K}$ state). Models are
needed to see if such a pattern is natural.
As the states that survive this cut appear to have
an affinity for S-wave, this may be evidence for $qq\bar{q}\bar{q}$
or $q\bar{q}q\bar{q}$ (as for example the $f_0(980)$) or for $gg$ content
(as perhaps in the case of $f_0(1500;1710)$ and $f_2(1930)$). It would be interesting
to study the production of known $q\bar{q}$ states in $e^+ e^- \to e^+ M e^-$
to see how they respond to this kinematic filter, and gain
possible insights into its dynamics.

Following this discovery there has been an intensive experimental
programme by the WA102 collaboration at CERN, which
has produced a large and detailed set of data on both the
$dP_T$
~\cite{ck97} and the azimuthal angle,
$\phi$,
dependence of meson
production (where
$\phi$ is the angle between the transverse momentum
vectors, $p_T$, of the two outgoing protons).
\par
The azimuthal dependences
as a function of
$J^{PC}$ and the momentum transferred at the proton vertices, $t$,
are also very striking. As described in ref.\cite{fecrome} here again the scalar mesons appear to divide into
 two classes: $f_0(980); f_0(1500);f_0(1710)$ which are all strongly peaked at small $\phi$ and
 the $f_0(1370)$ at large $\phi$. Exactly what this phenomenon implies for the dynamics and
 structure of these scalar mesons remains to be solved.

 One expects that there will be considerable mixing between the quenched glueball and the scalar mesons
 that were seeded by quarks. From a study of the $0^-0^-$ decays of the $f_0(1370;1500;1710)$, ref.
 \cite{CK02a} conclude that

 $\begin{array}{cccc}
 meson     & f_G & f_s & f_n \\ \nonumber
 f_0(1710) & 0.39(0.03) & 0.91 (0.02) & 0.15(0.02) \\ 
 f_0(1500) & -0.65(0.04) & 0.33(0.04)  & -0.70(0.07) \\ 
 f_0(1370) &  -0.69(0.07)& 0.15(0.01) & 0.70(0.07) \\ 
 \end{array}
$

\noindent This also intuitively is in line with the idea that $G$ and \ss ~mix to give the $f_0(1710)$;
that $G$ and \nn ~mix to give the $f_0(1370)$ and that that
 $G$, \ss ~and \nn ~mix to give the $f_0(1500)$ with a negative phase between the \nn-\ss ~(``flavour octet tendency").
 A new generation of experiments may enable flavour filtering in this extended nonet.

\subsection{Scalar mesons and glueballs}

Ref\cite{ClDK01} has shown that radiative transitions from excited
vector mesons to the scalar sector may be experimentally
accessible. In the simplest approach it assumed
  that there is no mixing among the scalars, so
that the $f_0(1370)$ is pure $n\bar n$ and the $f_0(1710)$ is pure $s\bar s$.

The $\rho_D(1700) \to \gamma f_1(1285) \sim 1MeV$ is a benchmark for experiments
to find (the $f_1$ is rather narrow which enables it to
be seen in $4\pi$ final states). If this can be verified then they should
look for the $\gamma f_0(n\bar{n}) \sim 0.9MeV$.
If the $f_0(1370)$ is entirely $n\bar{n}$ then the 0.9MeV will go entirely into it.
However, we anticipate glueball mixing into the 1370 (and other scalar mesons). If the glueball is light
($\sim 1300MeV$) it will mix strongly into the 1370 and dilute this 0.9MeV (it will be pushed into
the other scalars, e.g. the 1500). Conversely, if the glueball
is massive (up at 1700 MeV like Weingarten has argued) then the 1370 will remain rather pure $n\bar{n}$
and have a healthy radiative strength.

Ref.\cite{ClDK01} studied the effect of glueball mixing into the scalars and found that the
radiative transitions are potentially sensitive measures of this.
The result of the mixing is that the bare $n\bar n$ and $s\bar s$ states
contribute in varying degrees to each of the $f_0(1370)$, $f_0(1500)$, and
$f_0(1710)$.

Three different mixing scenarios have been proposed: the bare glueball is
lighter than the bare $n\bar n$ state (the light glueball solution); the
mass of the bare glueball is between the bare $n\bar n$ state and
the bare $s\bar s$ state (the middleweight glueball solution); and the mass
of the bare glueball is greater than the mass of the bare $s\bar s$ state.
The first two solutions have been obtained in \cite{Ba93,fcteper,CK02a} and the third
has been suggested in \cite{Se95,Wein}. The effects of the mixing on the
radiative decay widths of the $\rho(1700)$ and the $\phi(1900)$ to the
three $f_0$ states are given in ref.\cite{ClDK01} for each of these three cases.
The relative rates of the radiative decays of the $\rho(1700)$ to $f_0(1370)$
and $f_0(1500)$ change radically according to the presence
of the glueball admixture. So for a light glueball the decay to $f_0(1370)$
is relatively suppressed whereas for a heavy glueball it is substantial.
By contrast the effect on the decay to $f_0(1500)$ goes the other way.
Further, the $\phi(1900)$ would give a large width for the decay to
$f_0(1500)$ for a heavy glueball, but essentially zero for a light one.
The $f_0(1710)$ will be prominent in the decays of the $\phi(1900)$ for
all but the heaviest glueball.
It is clear that these decays do provide an effective
flavour-filtering mechanism.

Further, identifying the appropriate mixing
scheme gives insight into the underlying physics of glueballs.
The existing phenomenology from hadronic decays seems to favour a light
glueball. Essentially, if the decays of the ``bare'' glueball are
flavour-independent, then the observed flavour dependencies for the
hadronic decays of the physical mesons require \cite{CK02a} the glueball
mass to be at the
low end of the range preferred by quenched-lattice studies.

The resolution of the isoscalar-scalar problem is intimately connected
with the isovector-scalar problem. The existence of any $a_0$ other than
the $a_0(980)$ remains controversial. The different mixing schemes for the
isoscalar-scalar
mesons give rather different values for the mass of the bare $n\bar n$ state.
This mass will be reflected in the mass of its isovector partner, the $a_0$.
The width for the decay $\omega(1650) \to a_0\gamma$
is predicted to be large for a $q\bar{q}$ $a_0$, and is anticipated to be a well-defined decay. So this decay
can provide independent information on the existence and properties of
the $a_0(1450)$.
Predicting the width for $\omega \to \gamma a_0(980)$ is now also a challenge.

\section{Two coupled linear sigma models for two scalar nonets}

As we have seen in the previous discussion there seems to be a
proliferation of light scalar mesons. And as we have shown there
is good evidence that the spectrum below 1.7 GeV includes two light nonets of scalar
mesons, the heavier of which is the $q\bar q$ nonet
expected from QCD or the quark model, while the lighter is of more
complicated structure, but also in a nonet. The glueball of lattice gauge theory is mixed into
(at least some of) these states.

In order to have a realistic effective model at low energies for
the scalars we need an effective  chiral quark model, which includes all
scalars and pseudoscalars, and where the chiral symmetry is broken
by the vacuum expectation values of the scalar fields.

 The simplest such chiral quark model is the $U(N_f)_L\times
U(N_f)_R$ linear sigma model. Put as usual a scalar nonet into the
hermitian part of a $3\times 3$ matrix $\Phi$ and the associated
pseudoscalar nonet into the antihermitian part. For two scalar
nonets we need  another such $3\times 3$ matrix $\hat \Phi$. Let
the scalar \qq\ states above 1 GeV be in $\Phi$, while those below
1 GeV are in $\hat\Phi$. (In the approach of the unitarized quark
models discussed previously the states $\hat\Phi $ can be
generated by the unitarization.)

Then model both  $\Phi$ and $\hat\Phi$ by a gauged linear sigma
model, but with different sets of parameters ($\mu^2, \lambda$ ...
) and ($\hat\mu^2, \hat\lambda$ ...).

$$
 {\cal L}(\Phi ) =
\frac 1 2 {\rm Tr} [D_\mu\Phi D_\mu\Phi^\dagger] +\frac 1 2
\mu^2{\rm Tr} [\Phi \Phi^\dagger] -\lambda {\rm
Tr}[\Phi\Phi^\dagger \Phi\Phi^\dagger]\ + ... $$
$$
 \hat{\cal L}(\hat\Phi ) =
\frac 1 2 {\rm Tr} [D_\mu\hat \Phi D_\mu\hat\Phi^\dagger] +\frac 1
2 \hat\mu^2{\rm Tr} [\hat\Phi \hat\Phi^\dagger] -\hat\lambda {\rm
Tr}[\hat \Phi\hat\Phi^\dagger \hat\Phi\hat \Phi^\dagger]\ + ...
$$
 One can add further terms, but these are
not important for the present qualitative discussion. We have thus
doubled the spectrum and initially we have  two scalar, and two
pseudoscalar multiplets, altogether 36 states for three flavours.
We have not above included any flavour symmetry breaking nor the
glueball $G$ or anomaly terms for simplicity. The glueball would
mix with the singlet scalars through  terms like
$G$Tr$(\Phi+\Phi^\dagger)$.

 Then it is natural to introduce a coupling between the
two sets of multiplets, which can break the relative
symmetry\cite{nils0201171,fut}. The full effective Lagrangian for
both $\Phi$ and $\hat\Phi$ thus becomes,
$$ {\cal L}_{tot}(\Phi
,\hat\Phi ) ={\cal L}(\Phi )+\hat{\cal L} (\hat\Phi)+ \frac
{\epsilon^2} 4 {\rm Tr}[\Phi\hat\Phi^\dagger +h.c.]$$
   A similar scheme was discussed recently by Black et
al.\cite{black}, who also emphasized, with explicit examples, that
four-quark states of both the meson-meson type and of the Jaffe
type, $(\bar q\bar q)_3(qq)_{\bar 3}$, can be constructed, which
transforms in the same way under $SU(3)_L\times SU(3)_R$ as is
usually assumed for $q\bar q$ within chiral models. E.g., for a
lefthanded and righthanded diquark in the antisymmetric  $\bar 3$
representations of both colour and flavour one has

$$L^{cC}=\epsilon^{cab}\epsilon^{CAB}q^T_{aA}C^{-1}\frac{1+\gamma_5}{2}q_{bB},$$
$$R^{cC}=\epsilon^{cab}\epsilon^{CAB}q^T_{aA}C^{-1}\frac{1-\gamma_5}{2}q_{bB}.$$

Then the matrix $\hat \Phi_c^d=(L^{cA})^\dagger R^{dA}$ transforms
under $SU(3)_L\times SU(3)_R$ the same way as  a conventional
chiral $q \bar q$ state ($\Phi$).  Thus one may assume that $\Phi$
includes the $q\bar q$ nonets of scalars and pseudoscalars, while
$\hat \Phi$ stands for two extra chiral nonets of 4-quark states.

Now as a crucial assumption, let both $\Phi$ and $\hat\Phi$ have
vacuum expectation values (VEV)
 $v$ and $\hat v$ even if $\epsilon=0$ ($v=\mu^2/(4\lambda)+{\cal O}(\epsilon^2)$,
 $\hat v=\hat \mu^2/(4\hat\lambda)+{\cal O}(\epsilon^2)$.
The simplest physical intepretation of these VEV's is that
$v\propto <q\bar q>$ and $\hat v \propto <qq\bar q\bar q>$. With a
glueball one would expect that these would also have a gluonium
component.

The two originally massless pseudoscalar nonets then mix through
the $\epsilon^2$ term, with a mixing angle $\tan \theta=v/\hat v$,
such that one nonet remains massless while the other nonet obtains
a mass $m_\pi^2=\epsilon^2(v^2+\hat v^2)/v\hat v $. The mixing
angle is determined entirely by the two vacuum expectation values,
and is large if $v$ and $\hat v$ are of similar magnitudes,
independently of how small $\epsilon^2$ is as long as it stays
finite.

On the other hand the scalar masses and mixings  are only slightly
affected if $\epsilon^2/(\mu^2-\hat\mu^2)$ is small. They are
still close in mass to $\sqrt 2 \mu $ and $\sqrt 2 \hat \mu$ as in
the uncoupled case. These would be the two  scalar nonets we have
discussed in the previous chapters.

In order that this should have anything to do with reality, one
must of course get rid of the massless Goldstones.
 By gauging the overall axial symmetry
($D_\mu\Phi=\partial_\mu-ig[\lambda_iA_i\Phi+\Phi\lambda_iA_i ] $)
 the Higgs mechanism absorbs the
massless modes  from the model, but these degrees of freedom enter
instead as longitudinal axial vector mesons and give these mesons
(an extra) mass $m^2_A=2g^2 (v^2+\hat v^2)$. This is similar to
the original Yang-Mills theory and the work of Bando et
al.\cite{bando} on hidden local symmetries. Then with $\epsilon^2$
proportional to the average chiral quark mass one can interpret
the massive pseudoscalar nonet $\pi_a$ as the physical light
pseudoscalars. They would  be mixtures of the two original
pseudoscalar multiplets with a mixing angle $\theta$.

 The main prediction of this scheme is that one must
have doubled the light scalar meson spectrum, as seems to be
experimentally the case. Some more details are given
in\cite{fut}). Of course in order to make any detailed comparison
with experiment one must also break the flavour symmetry, and
unitarize the model. Especially the latter is not a simple matter
since it is a strong coupling model, although in principle
renormalizable.

The schizophrenic role of the pions in conventional models, as being
at the same time both Goldstone bosons and \qq\
pseudoscalars, is here resolved in a particularily simple way: One
has originally two Goldstone-like pions, out of which only one
remains in the spectrum, and which is a particular linear
combination of the two original pseudoscalar fields.

Both of the two scalar multiplets remain as physical states
 and one of these (formed by the $\sigma(600)$ and the $ a_0(980)$ in the case of  two flavours),
 or the $\sigma,\ a_0(980),\ f_0(980)$ and the $ \kappa\  $ in the case of three flavours
can then be looked upon as effectively a Higgs multiplet of strong
nonperturbative interactions when a hidden local  symmetry is
spontaneously broken.  The heavier scalar multiplet then being the
\qq\ scalars augmented by the glueball.

\section{Experimental Prospects}

Establishing that gluonic degrees of freedom are being excited is now
a real possibility.

 The scalars below 1 GeV are too light to enable a simple
distinction between loose molecules and compact four-quarks states to
be felt in decays (except perhaps for the $a_0(980)$ where the $K\bar{K}$
and $\eta\pi$ both couple strongly and point to a significant compact
four-quark feature). However, the production dynamics and systematics of
these states is interesting and full of enigmas, which may be soluble if
one adopts the four-quark/meson-meson bound state picture.

$D_s$ decays into $\pi f_0(980)$ clearly point to an $s\bar{s}$ presence in
the $f_0(980)$. However, the production in Z decays is rather
non-strange-like\cite{lafferty}.
$\psi$ to $\omega f_0$ and $\phi f_0$ also points towards
$n\bar{n}$ and $s\bar{s}$ structure in $f_0(980)$ and $a_0(980)$.
 The central production in pp shows that $f_0$ is strongly
produced, akin to other $n\bar{n}$ states and much stronger than
$s\bar{s}$ which appear to be suppressed in this mechanism. Furthermore,
$f_0$
survives the $dk_T \to 0$ filter of ref.\cite{ck97}.
 The systematics of this
appear to be driven by S-wave
production: this would be fine for either a compact four-quark or
molecule. We noted that there is also evidence for strong mixing between $f_0-a_0$ associated
with the nearby $K\bar{K}$ threshold. These phenomena fit more naturally with a
$qq\bar{q}\bar{q}$/meson-meson attraction as the controlling dynamics.

Whenever S-wave dynamics can play a role it will overide P-waves;
so one expects $K\bar{K}$ S-wave production to drive the $f_0/a_0$
whenever allowed. This is indeed what happens in the $\phi \to
\gamma f_0/\gamma a_0$; the``large" rate cries out for the
$K^+K^-$ loop to drive it. A question is whether the $s\bar{s}
n\bar{n}$ constituents of the intermediate state``between" the
initial $\phi$ and final $f_0$ are able to fluctuate spatially
enough to be identified as two colour singlet K's, which then
couple to the $f_0$, or whether they are a compact system in the
sense of being confined within $\sim$ 1fm. The former would have
some form factor suppression of the rate; the latter would be more
pointlike and larger rate. The emerging data are between these
extremes, but nearer to the expectations for a compact
$qq\bar{q}\bar{q}$ configuration. The \KK\ would then be the
long-range ($\geq O(\Lambda_{QCD})$) tail.

Knowledge on the $\gamma \gamma$ couplings is lacking and better
data would be useful; however, it is not immediately clear how
this probes the deep structure of the scalar mesons. We know that
for the $2^{++}$ $\gamma \gamma$ reads the compact $q\bar{q}$
flavours; there is no 2-body S-wave competition in the imaginary
part as $\rho\rho$ etc are too heavy. One would expect that for
the $0^{++}$ the $K\bar{K}$ will dominate the $\gamma \gamma$ if
there is a long-range \KK\ component in the wavefunction. At the
other extreme; were the state a pure compact four quark, then
higher intermediate states - KK, KK*,KK** etc - would all be
present. Achasov\cite{achasov3} has discussed these and a precise
calculation has many problems, but the {\it ratios} of $\gamma
\gamma$ to $f_0/a_0$ would probably be sensitive and more
reliable.

The production by highly virtual $\gamma^*\gamma^*$ in $e^+e^- \to e^+e^- f_0/a_0$
could probe the spatial dependence of their wavefunctions. It would be especially instructive
were the ratio to be strongly $Q^2$ dependent.

In summary, the theoretical frontier suggests that
 one can divide the phenomenolgy of scalars into those above and
those below 1 GeV. We suspect that much of the confusion begins to
evaporate if one adopts such a starting point. Empirically, signs
of gluonic excitation are appearing: (i) in the form of hybrids
with the exotic $J^{PC} = 1^{-+}$ now seen in various channels and
more than one experiment; (ii) with $0^{-+}$ and $1^{--}$ signals
in the 1.4 - 1.9 GeV region that do not fit well with conventional
quarkonia and show features predicted for hybrids; (iii) in the
form of the scalar glueball mixed in with quarkonia in the 1.3 -
1.7 Gev mass range. Theoretical questions about the latter are
concerned with whether the effects of the glueball are localised
above 1 GeV, or whether they are spread across a wider mass range,
even down to threshold. Experimental questions that need to be
resolved concern the existence and properties of the $f_0(1370)$
and $a_0(1450)$.

These questions in turn provoke a list of challenges for experiment.

(i) In $e^+e^-$, or vector meson photoproduction at Fermilab and
Jefferson Laboratory
high statistics studies of
radiative decays of such states into the $f_0(980);f_1(1285);
f_2(1270)$ could teach us much\cite{ClDK01}.

(ii) $\gamma \gamma$ couplings give rather direct information on the
flavour content of C=+ states. Such information on the scalar mesons
will be an essential part of interpreting these states.
The $Q^2$ dependence of $\gamma^*\gamma^*$ in $e^+e^- \to e^+e^- f_0/a_0$
could probe the spatial dependence of these states. Complementing this $ep \to ep 0^{++}$
could probe their $\gamma^* \omega$ couplings.

(iii) Heavy flavour decays, in particular $D_s$ and $D$ into $\pi$ and associated
hadrons can access the scalar states. Precision data are needed to disentangle the
contributions of the various diagrams, whereby the flavour content of the scalars
can be inferred. There is also a tantalising degeneracy between the $\pi_g(1.8)$ and
the $D$, which may radically affect the Cabibbo suppressed decays of the latter.
Hence precision data on such charm decays is warranted for reasons that go far beyond
simply issues about scalar mesons.

(iv) Tau Charm Factories may at last appear.
 $\chi$ decays offer an entree into light flavoured states; the
excitement about the scalar glueball mixing with the quarkonia
nonet began when the precision data from $p\bar{p}$ annihilation
at LEAR first emerged. Data at rest were beautiful and well
analysed. Data in flight however tend to be more problematic, not
least as one cannot so easily control knowledge of the incident
partial wave. $\chi$ decays can access these phenomena, at c.m.
energies up to 3.5 GeV, and from well defined initial $J^{PC}$
states. $\chi_2 \to f_2 +(\pi\pi;K\bar{K};\eta\eta)$ will favour
the $0^{++}$ channel and flavour select \nn\ and \ss\ components
if the $f_2(1270)$ or $f_2(1525)$ is selected respectively. The
$1^+$ nonet is not flavour ideal\cite{Cl97} but $f_1(1285)$ is
narrow and potentially a clear signal against which the scalar
hadrons can also be formed in S-wave. (The $\chi_1 \to \pi +
1^{-+}$ also provides an entree into the exotic hybrid channel)

\section{Conclusions}

There seems to be growing experimental
evidence for two light nonets of scalars, one in the 500-1000 MeV
region ($\sigma(600),$ $ a_0(980),$ $f_0(980),$ $\kappa (?)$), and
another one near 1.3-1.7  GeV ($f_0(1370),$ $
f_0(1500)/f_0(1710),$ $ a_0(1450),$ $ K^*_0(1430)$) where the
``overpopulation" and systematics of the latter in particular fit
with the notion that the scalar glueball of lattice QCD is mixing
with these states. A linear sigma model has been proposed as a
``toy model" for each multiplet, each one with its separate
 vacuum expectation value. Then after coupling these two models
through a mixing term and gauging the overall symmetry, one can argue that one of the nonets
(the lighter one) is a true Higgs nonet for strong interactions and the heavier one is strongly influenced by
mixing with the scalar glueball.

A more detailed understanding requires a unitarized model whereby one can
undertand how the masses are shifted, the widths and mixings are  distorted, and even why new
scalar meson-meson resonances can be created in nonexotic
channels. These effects are
important  because of the large effective coupling, and because of the nonlinearities due to the S-wave cusps.

Finally, we note that no anomalies are anticipated, nor seen, for
scalar mesons made from heavy flavours. While this remark may appear to be trivial, it
reinforces the singular properties being manifested for the light scalars. Given that QCD effects,
 such as a glueball
and strong attraction in S-waves are naturally expected in the light mass region, one may qualitatively conclude
that these singular properties are in accord with theoretical prejudices. Isolating their dynamics in detail may therefore shed
new light on the nature of confinement in QCD and of effective theories incorporating Higgs' ideas.

\begin{center}
{\bf Acknowledgements}
\end{center}
\par

 Support  from EU-TMR program ``EURODAFNE", contract
CT98-0169 is gratefully acknowledged.


\begin{thebibliography}{99}
\bibitem{nils0201171} N. A. T\"ornqvist, (talk at IPN Orsay
workshop on "Chiral fluctuations in hadronic matter",   Sept.
26-28. 2001), Paris, France, Proceedings edited by Z. Aouissat et
al., hep-ph/0201171.
\bibitem{Ba97} G. Bali et al., SESAM Collaboration; {\it Nucl.Phys.Proc.Suppl.} {\bf 63}
 209 (1997).
\bibitem{Ba93} G. Bali et al., {\it Phys.Lett.} B{\bf 309} 378
(1993).
\bibitem{Se95} J. Sexton et al., (IBM Collaboration) {\it Phys. Rev.Lett.} {\bf 75} 4563
(1995).
\bibitem{fcteper} F.E. Close and M.J. Teper, ``On the lightest Scalar Glueball"
Rutherford Appleton Laboratory report no. RAL-96-040; Oxford
University report no. OUTP-96-35P.
\bibitem{Mo97} C.J. Morningstar and M. Peardon, {\it Phys.Rev.} D {\bf 56} 4043
(1997).
\bibitem{Wein}
D. Weingarten, {\it Nucl. Phys. Proc. Suppl.} {\bf 53} 232 (1997);
{\bf 63} 194 (1998); {\bf 73} 249 (1999).
\bibitem{Jaf00}
M. Alford and R.L. Jaffe, {\it Nucl. Phys.} {\bf B578} 367 (2000),
(hep-lat/0001023).
\bibitem{Ja77} R.L. Jaffe, {\it Phys.Rev.} D {\bf 15} 281 (1977);
 R.L.Jaffe and F.E. Low, {\it Phys.Rev.} D {\bf 19} 2105 (1979).
\bibitem{Ja00} R.L. Jaffe, hep-ph/0001123.
\bibitem{12gev} {\it The Science Driving the 12 GeV Upgrade of CEBAF} Jefferson Lab
report 2001 (eds L. Cardman, R. Ent, N. Isgur et al.), pp.19-22.
\bibitem{isgur1} S. Godfrey and N. Isgur, {\it Phys.Rev.} D {\bf 32} 189
(1985).
 \bibitem{Go99} S. Godfrey and J. Napolitano, {\it Rev. Mod. Phys.} {\bf 71} 1411
 (1999).
\bibitem{Am95}
C. Amsler and F.E. Close, {\it Phys. Lett.} B {\bf 353 } \rm 385
(1995).
\bibitem{An98} V. V. Anisovich, {\it Physics-Uspekhi} {\bf 41} 419
(1998).
\bibitem{McN00} C. McNeile and C. Michael, {\it Phys.Rev.} D {\bf 63} 114503 (2001). (hep-lat/0010019).
\bibitem{ClK01a} F.E. Close and A. Kirk, {\it Phys.Lett.} B {\bf 483} 345
(2000).
\bibitem{klempt} E. Klempt, PSI Proceedings,00-01 61 (2000).
\bibitem{FC00}
F.E.Close, {\it Acta Physica Polonica} {\bf B31} 2557 (2000).
\bibitem{minkochs} P. Minkowski and W. Ochs, {\it Eur.Phys.} J {\bf C9} 283
(1999).
\bibitem{kappadata} E. M. Aitala et al. (E791 Collaboration) hep-ex/0204018;
C. G\"obel, (E791 Collaboration) hep-ex/0012009; S. Kopp et al.,
(CLEO Collaboration) {\it Phys.Rev.} D {\bf 63}. 092001 (2001).
\bibitem{kappa} M. R. Pennington hep-ph/9905241.
\bibitem{kappaphen}  S. Spanier and N.A. T\"ornqvist,``Note on scalar
mesons" in D.E. Groom et al., [Particle Data Group], {\it
Eur.Phys.J.} {\bf C15} 1 (2000).
\bibitem{iswein} J. Weinstein and N. Isgur, {\it Phys. Rev.Lett.} {\bf 48} 659 (1982);
{\it Phys. Rev.} D {\bf 27} 588 (1983).
\bibitem{tornqv} N.A. T\"ornqvist, {\it Zeit. Phys.} C {\bf 68} 647 (1995).
\bibitem{penn} M. Boglione and M.R.Pennington, {\it Phys. Rev.Lett.} {\bf 79} 1998
(1997).
\bibitem{achasov2}
N.N. Achasov and G.N. Shestakov, {\it Phys. Rev.} D {\bf 56} 212
(1997).
\bibitem{achasov1}
N.N. Achasov, S.A. Devyanin and G.N. Shestakov, {\it Phys. Lett.}
B {\bf 88} 367 (1979).
\bibitem{speth}
O. Krehl, R. Rapp and J. Speth, {\it Phys. Lett.} {\bf B390} 23
(1997).
\bibitem{ClK01b} F.E. Close and A. Kirk, {\it Phys. Lett.} B {\bf 489} 24
(2000).
\bibitem{ClK01c} F.E. Close and A. Kirk, hep-ph/0106108, {\it Phys. Lett.} B
{\bf 515} 13
(2001).
\bibitem{kloe} A. Aloisio et al., (KLOE Collaboration)
hep-ex/0107024.
\bibitem{kloe2} A. Aloisio et al., (KLOE Collaboration)
hep-ex/0204012;  and hep-ex/0204013.
\bibitem{jaffe01} R.L. Jaffe (private communication).
\bibitem{dunwoodie}
W. Dunwoodie, Proceedings of Hadron 97, AIP Conf. Series {\bf 432}
p.753 (1997).
\bibitem{e687} P. Frabetti et al., (E687 Collaboration) {\it Phys.Lett.} B {\bf 351}
591 (1995).
\bibitem{e791}
E Aitala et al., (E791 Collaboration) {\it Phys.Rev.Lett.} {\bf
86} 765 (2001).
\bibitem{pdg2000} D.E. Groom et al., {\it Eur. J. Phys.} {\bf C 15}, 1 (2000).
\bibitem{kyoto} Conference: ``Possible existence of the light $\sigma$
resonance and its implications to hadron physics", Kyoto, Japan
11-14th June 2000, KEK-proceedings/2000-4; See also N.A.
T\"ornqvist, ``Summary of the conference", hep-ph/0008135.
\bibitem{wu} N. Wu, hep-ex/0104050.
\bibitem{barnes02} T. Barnes, Hadron 2001 Confrerence Summary: Theory; hep-ph/0202157.
\bibitem{aston} D. Aston et al., {\it Nucl. Phys.} {\bf B296} 493 (1988).
\bibitem{au} K.L. Au, D. Morgan and M.R. Pennington, {\it Phys. Rev.} {\bf D 35}, 1633
(1987).
\bibitem{CLEO} D.M. Asner  et al., (CLEO collaboration), {\it Phys.
Rev.} {\bf D61}, 0120002 (1999).
\bibitem{wein90} S. Weinberg, {\it Phys. Rev. Lett.} {\bf 65} 1177 (1990).
\bibitem{gatto} R. Gatto, G. Nardulli, A.D. Polosa and N.A. T\"ornqvist,
 {\it Phys. Lett.} {\bf B494} 168 (2000).
\bibitem{rass} A.D. Polosa, {\it The CQM model}, Nuovo Cimento {\bf 23} 1 (2000).
\bibitem{WSB} M. Bauer, B. Stech and M. Wirbel, {\it Z. Phys.} {\bf
C16}, 205 (1983).
\bibitem{volkoff} D. Ebert and M.K. Volkov, {\it Z. Phys.} {\bf C16}, 205
(1983).
\bibitem{paver} N. Paver and Riazuddin, hep-ph/0107330.
\bibitem{deandrea} A. Deandrea, R. Gatto, G. Nardulli, A.D. Polosa and N.A. T\"ornqvist,
{\it Phys. Lett.} {\bf B502} 79 (2001).
\bibitem{lafferty} G. Lafferty, (private communication).
\bibitem{delphi} M. Chapkin et al., {\it DELPHI 2001-063 CONF491}, Lepton Photon Conf, Rome, 2001.
\bibitem{ach81} N.Achasov, S.Devyanina and G.Shestakov, Sov.J.Nuc.Phys. {\bf 31}, 715
(1981).
\bibitem{fecrome} F.E. Close, Rapporteur talk at Lepton Photon Conf, Rome, 2001
hep-ph/0110081.
\bibitem{sumpap} A. Kirk, {\it Phys. Lett.} B {\bf 489} 29 (2000).
\bibitem{cik}
F.E. Close, N. Isgur and S. Kumano, {\it Nucl.Phys.} B {\bf 389}
513 (1993).
\bibitem{achasov3}
N. Achasov and V. Ivanchenko, {\it Nucl.Phys.} B {\bf 315} 465
(1989).
\bibitem{cbrown}  N. Brown and F.E. Close, in {\it The DA$\Phi$NE Physics Handbook},
L. Maiani, G. Pancheri and N. Paver, eds.; INFN Frascati (1995).
\bibitem{lucieu} J.L. Lucio and J.M. Pestieau, {\it Phys. Rev.} D {\bf 42} 3253
(1990).
\bibitem{achasov4} N.N. Achasov and V.V. Gubin, {\it Phys. Rev.} {\bf
D63} 094007 (2001)  (hep-ph/0101024).
\bibitem{ach2001} N.N.Achasov and A.Kiselev, {\it Phys. Lett.} {\bf B534} 83
(2002) (hep-ph/0203142)
\bibitem{barnes} T. Barnes, {\it Phys.Lett.} B {\bf 165} 434
(1985).
\bibitem{u3u3} J. Schechter and Y. Ueda, {\it Phys. Rev.} {\bf D3}, 2874 (1971).
\bibitem{lsm} N.A. T\"ornqvist, {\it Eur. J. Phys.} {\bf C11}, 359 (1999);
M. Napsusciale, hep-ph/9803396; see also G. Parisi and M. Testa
{\it Nuov. Cim.} {\bf LXVII}, 13 (1969).
\bibitem{NAT1} N.A. T\"ornqvist and M. Roos, {\it Phys. Rev. Lett.} {\bf 76}, 1575
(1996).
\bibitem{morgan} D. Morgan and M.R. Pennington, {\it Phys. Rev.} {\bf D48},
1185; (1993); {\it ibid.} {\bf D48}, 5422 (1993).
\bibitem{bogl} M. Boglione and M. R. Pennington, hep-ph/0203149.
\bibitem{cherry} S. N. Cherry and M. Pennington, Nucl. Phys {\bf A688} 823 (2001).
\bibitem{beve} E. Van Beveren et al., {\it Z. Physik} {\bf C30}
615 (1986).
\bibitem{black} D. Black, A. Fariborz, S. Moussa, S. Nasri, J. Schechter,
{\it Phys. Rev.} {\bf D64}, 014031 (2001).
\bibitem{modelskappa}
 S. Ishida et al., {\it Prog.
Theor. Phys.} {\bf 98} 1005 (1997) {\it ibid} {\bf 101} 661
(1999); D. Black et al., {\it Phys. Rev.} {\bf D58} 054012 (1998);
{\it ibid} {\bf 59} 074026 (1999); J.A. Oller, E. Oset, J.R.
Pelaez, Phys.Rev.{\bf D59} 074001 (1999); {\it Erratum-ibid.} {\bf
D60} 099906 (1999); J.A. Oller and E. Oset, {\it Phys. Rev.} {\bf
D 60} 074023 (1999); {\it ibid} {\bf 60} 099096 (1999).
\bibitem{flatte} S. M. Flatt\'{e}, {\it Phys. Lett.} {\bf 63B} 224 (1976).
\bibitem{chan} M. Chanowitz, {\it Phys. Rev.Lett.} {\bf 46} 981
(1981).
\bibitem{fcrpp} F.E. Close, {\it Rep. Prog. Phys.} {\bf 51} 833 (1988).
\bibitem{robson} D. Robson, {\it Nucl. Phys.} B { \bf 130} 328 (1977).
\bibitem{cfl} F.E. Close, G. Farrar and Z.P. Li, {\it Phys. Rev.} D {\bf 55} (1997)
5749.
\bibitem{ck97} F.E. Close and A. Kirk, {\it Phys. Lett.} B {\bf 397} 333 (1997).
\bibitem{Cl97} F.E. Close and A. Kirk, {\it Z.Phys.} C {\bf 76} 469
(1997).
\bibitem{CK02a} F.E. Close and A. Kirk, {\it Z.Phys.}. {\bf C21} 531
(2001).
\bibitem{ClDK01} F. E, Close, A.Donnachie and Yu Kalashnikova
(hep-ph/0201043).
\bibitem{dib} C. Dib and R. Rosenfeld, {\it Phys. Rev.} {\bf 63} 11750 (2001).
\bibitem{fut} N.A. T\"ornqvist ``The lightest scalar nonet as  Higgs bosons of strong interactions"
 hep-ph/0204215.
\bibitem{bando} M. Bando, T. Kugo, S.Uehara, K.Yamawaki,T. Yanagida, {\it Phys. Rev. Lett.} {\bf 54}  1215 (1985).

\end{thebibliography}
\end{document}